# All You Need is Color: Image based Spatial Gene Expression Prediction using Neural Stain Learning

Muhammad Dawood[1][0000-0001-5358-9478], Kim Branson[2], Nasir M. Rajpoot[1][0000-0002-4706-1308], Fayyaz ul Amir Afsar Minhas[1][0000-0001-9129-1189]

[1]Department of Computer Science University of Warwick, Coventry, UK
[2]Artificial Intelligence & Machine Learning, GlaxoSmithKline

```
{Muhammad.Dawood, N.M.Rajpoot,
Fayyaz.Minhas}@warwick.ac.uk, kim.m.branson@gsk.com
```

**Abstract.** "*Is it possible to predict expression levels of different genes at a given spatial location in the routine histology image of a tumor section by modeling its stain absorption characteristics?*" In this work, we propose a "stain-aware" machine learning approach for prediction of spatial transcriptomic gene expression profiles using digital pathology image of a routine Hematoxylin & Eosin (H&E) histology section. Unlike recent deep learning methods which are used for gene expression prediction, our proposed approach termed Neural Stain Learning (NSL) explicitly models the association of stain absorption characteristics of the tissue with gene expression patterns in spatial transcriptomics by learning a problem-specific stain deconvolution matrix in an end-to-end manner. The proposed method with only 11 trainable weight parameters outperforms both classical regression models with cellular composition and morphological features as well as deep learning methods. We have found that the gene expression predictions from the proposed approach show higher correlations with true expression values obtained through sequencing for a larger set of genes in comparison to other approaches.

## 1 Introduction

Gene expression quantification plays a key role in understanding cancer genetics and identifying potential targets for novel therapeutics. For example, cancer driver mutations can be identified by comparative analysis of differential gene expression levels across normal and cancerous tissues, which can then be used for targeted therapy [1]–[4]. To determine gene expression in a tissue sample, transcriptomic sequencing technologies such as bulk RNA-Seq [5] and single-cell RNA-sequencing (scRNA-seq) [6], [7] are used. Bulk RNA-Seq gives expression of different genes across different cell types, whereas single-cell RNA-Seq estimates gene expression at cellular level.

However, both these technologies fail to capture the spatial variations in gene expression profile across a tissue sample, which is crucial when studying tumor heterogeneity [8].

To tackle this dilemma, one potential solution is Spatial Transcriptomics (ST). It is a relatively new technology that measures spatially resolved messenger RNA (mRNA) profile in a tissue section using unique DNA barcodes [9]. These unique DNA barcodes are used to map the expression of thousands of genes at each spot in a tissue slide. For each tissue section, ST methods generate a local spot-level gene expression profile together with a whole slide image (WSI) for the Hematoxylin & Eosin (H&E) stained tissue section. This information can be very beneficial in terms of understanding gene expression variations across different regions in a given tumor sample and has been used to gain valuable insights into the role of various different kinds of cells in the tumor microenvironment (TME) and their impact on response to therapy [10], [11].

An interesting question in this regard is whether, and to what degree, it is possible to predict gene expression profiles from the WSI of H&E stained tumor section alone using the ST data as ground-truth. This association of visual characteristics of the tissue with gene expression profiles can provide new insights into the local mapping of various different kind of cells in the TME and can lead to possible discovery of visual cues associated with expression profiles of different genes [12]. Deep learning has been used to predict genetic mutations [13]–[16], gene and RNA-Seq expression profile [17], Microsatellite Instability (MSI) [18], [19], and Tumor Mutation Burden (TMB) [20], [21] from WSIs of H&E sections. Since these methods have been developed using bulk RNA-Seq data, where the target gene expression profile is available at the tissue level and provides a coarse-grained phenotype only, these methods may not be able to uncover exact visual patterns that correlate with a specific gene expression profile. In contrast, the target label in the ST data is available at spot level which provides a more fine-grained local mapping of transcriptomic variation across the tumor tissue. Recently a deep learning method was proposed for predicting spatially resolved gene expression profile from spatial transcriptomic imaging data [22].

A vast majority of existing methods for computational pathology in general, and for image based prediction of local gene expression profiles in particular, use convolutional neural networks such as DenseNet [23], ResNet [24], often pretrained on natural images and finetuned on computational pathology tasks. Most of them fail to explicitly model tissue staining in their design, which is a fundamental aspect of computational pathology images. Pathology images are obtained by staining a given sample with a dye that absorbs incident light depending upon dye concentration and its binding characteristics for different components (e.g., proteins, DNA, etc.) in the sample. Routine stains such as Hematoxylin, Eosin and special stains for antibody-optimized immunohistochemical (IHC) markers are often used for pathology diagnosis and/or biomarker analysis. Consequently, the image acquisition process in routine histopathology and ST is predominantly based on light absorption. This contrasts with natural images obtained through standard digital cameras, such as those in the widely used ImageNet database, which are unstained and operate on a different lighting and camera model involving reflected, absorbed, and radiated light. Despite this fundamental difference, machine learning models in computational pathology and Spatial Transcriptomics Imaging are not explicitly "stain-aware". Stain normalization and separation [25]–[29] are used as pre-processing steps in computational pathology algorithms to control variations resulting from differences in slide preparation protocols and digital slide scanning with scanners from different vendors. In recent years, stain and color-based augmentation of images in deep learning models are also used to overcome such variations [30], [31].

In this work, we investigate the contribution of stain information contained in WSIs of routine H&E stained tissue sections with associated ST data for image based local gene expression prediction directly without using computationally intensive deep learning approaches. We propose a novel neural stain deconvolution layer that can model stain deconvolution in an end-to-end manner. We show that the proposed neural stain learning (NSL) can model the prediction of local gene expression and lead to statistically significant correlation scores between true and predicted expression levels for a larger number of genes in comparison to existing methods using deep convolutional networks or cellular composition or cellular morphology based regression. The simplicity of our NSL method is underscored by the fact that the number of trainable weight parameters in the proposed scheme is significantly smaller (11 per gene), as compared to millions of learnable parameters in the case of a convolutional network.

## 2 Materials and Methods

The workflow of the proposed approach for image based local gene expression prediction from WSI of routine H&E tumor section with associated ST data is shown in Fig. 1. We take an image patch corresponding to an ST spot as input and generate gene expression prediction from visual information contained in the image patch. The model is trained using WSIs and associated local gene expression data from ST experiments. Below we provide details about the dataset and pre-processing, problem modeling and model training and evaluation.

### 2.1 Dataset and Preprocessing

We used a publicly available dataset [32], consisting of 36 tissue histology slides acquired from eight Human Epidermal growth factor Receptor 2 (HER2) positive breast cancer patients. WSIs in this dataset were obtained by sectioning frozen tissue sample at 16μm, after staining with H&E and scanning at 20× objective. Each WSI contains multiple spots arranged in a grid like pattern for measurement of spatial gene expression profiles of 11,880 genes. On average, there are 378 spots per slide with a total of 13,620 spots in the dataset. Gene expression data was normalized using regularized negative binomial regression method. From each WSI, patches of size 256×256 pixels (corresponding to a tissue area of 170×170μm$^2$) were taken from the center of each spot, making the tissue region captured by the cropped image slightly larger than the actual spot size (100×100μm$^2$). Genes at each spot were filtered based on their median expression level and the top 250 genes with significant expression across all spots were analyzed. As the gene expression data distribution was highly skewed (some values were too high and some of them were zero), we transformed the gene expression data using a logarithmic transformation after addition of a pseudo-count value.

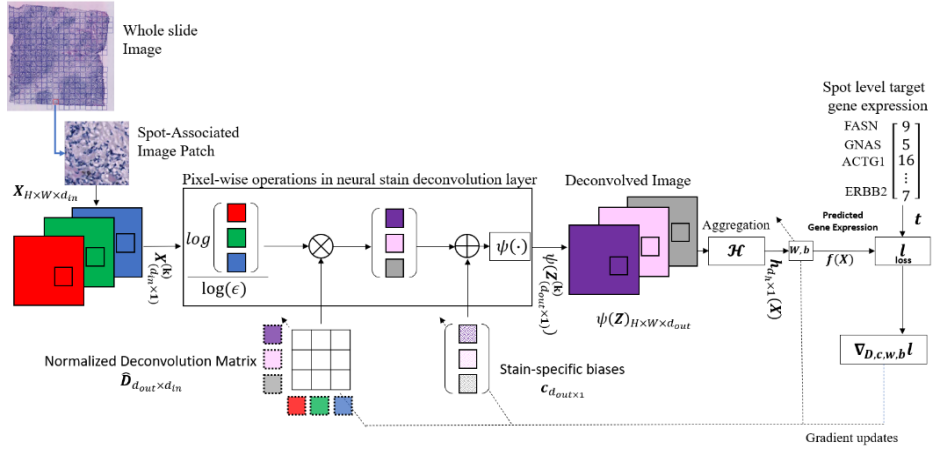

**Fig. 1.** Workflow of the proposed approach for gene expression prediction task. From each WSI, image patches corresponding to all ST spots are extracted using spatial coordinate data, and the spot level gene expression profile was used as a target label for learning problem-specific stain matrix. Deconvolved pixels values for the image patch are then aggregated to get a single patch-level feature for the gene expression prediction task.

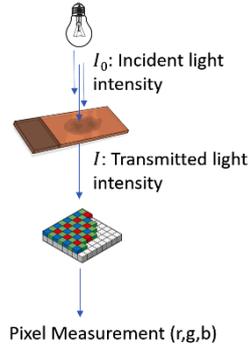

**Fig. 2.** Simplified diagram of whole slide image scanning of a stained histological sample.

### 2.2  Prediction of Spatial Gene Expression using neural stain learning

In order to model the problem of predicting gene expression levels of a number of genes at a given spot from the corresponding RGB image as a learning problem, consider a training dataset $\mathcal{B} = \{(\mathbf{X_i}, \mathbf{t_i}) | i = 1 \ldots N\}$ comprising of $N$ spots and their associated gene expression profiles. We denote the image patch corresponding to a spot by $\mathbf{X_i}$ and its associated target vector of normalized gene expression scores for a number $G$ of genes by $\mathbf{t_i} \in \Re^G$. The objective of the learning problem is to estimate learnable parameters $\boldsymbol{\theta}$ of a predictor $\mathbf{y}_i = \boldsymbol{f}(X_i; \boldsymbol{\theta})$ such that the output of the function $\boldsymbol{f}$ matches target gene expression levels for test images. Several recent methods for image based local gene expression prediction tend to obtain the prediction function using deep neural network models. In this work, we modelled this problem as a stain estimation problem based on the hypothesis that the gene expression level of a gene at a spot affects the degree of staining/dye absorption in the tissue leading to changes in pixel values in the corresponding image. Therefore, if we can estimate the association between stain variations in an image and the corresponding gene expression patterns of a gene using a training dataset, it would then be possible to infer gene expression levels of a gene for a test image. The underlying concept for this approach is shown in Fig. 2. The observed RGB intensity value at a pixel in the scanned image of a tissue sample at a given spot is dependent upon light absorption characteristics of the dye used to stain the tissue sample, the amount of stain or dye absorbed by the tissue sample as well as the intensity of incident light [25]. Based on the Beer-Lambert law, the vector of stain intensities $\mathbf{z}$ at a given pixel can be estimated from the corresponding RGB pixel values $\boldsymbol{x}$ by the relation $\mathbf{z} = -\boldsymbol{D}\log(\boldsymbol{x})$ where $\boldsymbol{D}$ is a stain deconvolution matrix which determines how different stains in the given image correspond to color pixel intensities [26]. The deconvolution matrix corresponding to standard stains such as Hematoxylin, Eosin and DAB are available in the literature along with a number of methods for estimating the prevalent stain vectors and normalizing stain variations in images [25]–[29].

In this work, our goal is to estimate the deconvolution matrix in such a way that the resulting stain intensities can explain variations in gene expression patterns. Fig. 1 shows the general framework of the proposed approach called Neural Stain Learning (NSL). NSL assumes a randomly initialized stain deconvolution matrix. It takes an input image corresponding to a spot and deconvolves each pixel value in the input image to generate a vector of stain intensities which form the deconvolved image. This stain deconvolved image is then used to regress the gene expression profile corresponding to a given spot. The gradient of the difference between predicted and target gene expression values in the training dataset is used to update the elements of the deconvolution matrix and parameters of the regressor in an end-to-end manner using gradient descent optimization.

More formally, assume that $\boldsymbol{X}^{(k)}$ is a column vector of RGB values corresponding to a single pixel $k = 1 \ldots K$ in input image $\boldsymbol{X}$ with each pixel value in the range $[\epsilon, 1]$ (with small $\epsilon > 0$). Also, assume that $\widehat{\boldsymbol{D}}$ is a row-normalized form of the randomly initialized de-convolution matrix $\boldsymbol{D}$, i.e., for each row $j$, $\widehat{\boldsymbol{D}}_j = \frac{D_j}{\|D_j\|}$. The corresponding stain intensities in the stain deconvolved image $\boldsymbol{Z}$ can thus be obtained as: $\boldsymbol{Z}^{(k)} = \widehat{\boldsymbol{D}} \frac{\log(X^k)}{\log(\epsilon)}$. As discussed earlier, the goal of NSL is to obtain an optimal "pseudo"-deconvolution matrix $\boldsymbol{D}^*$ that allows prediction of target values of the training images. We denote the overall prediction function by $\boldsymbol{f}(X; \boldsymbol{D}, \boldsymbol{\theta})$ which has two sets of learnable parameters – the deconvolution matrix $\boldsymbol{D}$ which is used to produce stain intensities at each pixel in the image and weight parameters $\boldsymbol{\theta}$ that are used to predict gene expression levels based on these stain intensities. The overall learning problem can be written as the following empirical risk minimization with the loss functional $l(\cdot,\cdot)$:

$$\mathbf{D}^*, \boldsymbol{\theta}^* = \underset{\mathbf{D}, \boldsymbol{\theta}}{\operatorname{argmin}} L(f; \mathcal{B}) = \sum_i l(f(X_i; \mathbf{D}, \boldsymbol{\theta}), t_i)$$

In this work, we have used the simple mean squared error loss function. As a minimal learning example and without loss of generality, we can assume a single downstream neuron which operates on aggregated stain intensities of the input image to predict the expression of a single gene. More specifically, the predictor can be written as follows:

$$f(X; \mathbf{D}, \boldsymbol{\theta} = (\mathbf{w}, b, \mathbf{c})) = \mathbf{w}\mathcal{H}_{k=1\ldots K}\left\{\psi\left(\widehat{\mathbf{D}}\frac{\log(X^{(k)})}{\log(\epsilon)} + \mathbf{c}\right)\right\} + b$$

where $\mathbf{c}$ is an (optional) 3x1 vector of "stain"-wise learnable bias parameters such that $c_j$ is added to the jth channel of $\mathbf{Z}^{(k)} = \widehat{\mathbf{D}}\frac{\log(X^k)}{\log(\epsilon)}$. $\psi(\cdot)$ is an activation function (bipolar sigmoid) that operates on the deconvolved stain output $\mathbf{Z}^{(k)}$ and $\mathcal{H}_{k=1\ldots K}$ is an operator that aggregates the transformed stain values via simple averaging:

$$\mathcal{H}_{k=1\ldots K}\{\psi(\mathbf{Z}^{(k)})\} = \frac{1}{K}\sum_{k=1}^{K}\mathbf{1}_3^T\psi(\mathbf{Z}^{(k)}).$$

The output of the aggregation is then fed into a single neuron with weights **w** and bias b which generates a prediction corresponding to a single gene. The proposed architecture can be implemented with any automatic differentiation package such as PyTorch or TensorFlow by computing the gradient $\nabla L(f; \mathcal{B})$ of the loss function with respect to all learnable parameters in the above model. At each optimization step, the deconvolution matrix is row-normalized to yield $\widehat{\mathbf{D}}$. It is important to note that $\psi\left(\widehat{\mathbf{D}}\frac{\log(X^{(k)})}{\log(\epsilon)} + \mathbf{c}\right)$ in the proposed model can be thought of as a general neural stain deconvolution layer with a single neuron with 3 inputs and 3 outputs whose weights constitute the normalized stain deconvolution matrix $\widehat{D}$. These weights are shared across the pixels of all training images. This results in a very small number of learnable weight parameters (11 per gene – 6 independent weights in the row-normalized $3 \times 3$ matrix $\widehat{\mathbf{D}}$, 3 optional stain-wise biases and a weight and bias parameter for the neuron used for generating the gene level output) for the overall prediction model which is much smaller than the millions of weights used in classical deep learning architectures. Furthermore, the proposed architecture is not specific to any particular type of learning problem and can be applied to other learning problems in absorption microscopy.

## 2.3   Model training and Evaluation

The generalization performance of the proposed model was evaluated using leave one patient out cross-validation, i.e., data for one patient was held out for testing and training was performed on the remaining patients. For the test patient, we calculated the Pearson correlation coefficient of the predicted and true/target gene expression (from the corresponding ST spot) with its associated *p*-value. In order to analyze the predictive accuracy of the proposed method, we used the median correlation score of a gene across all patients as a performance metric. The *p*-values associated with the correlation score of a given gene across multiple cross-validation runs were combined by calculating the median p-value ($p_{50}$) and using $2p_{50}$ as a conservative estimate for significance [33]. In each cross-validation run, the model was trained for 250 epochs using the adaptive momentum based optimizer [34] with a learning rate of 0.001 and batch size of 128.

### 2.4 Comparison with other approaches

We compared the performance of the proposed model with methods that use deep learning or cellular features for gene expression prediction. In this section, we describe details of experiments for comparative performance evaluation.

**Comparison with DenseNet121**

To compare our results with deep learning methods, we fine-tuned ImageNet [35] pretrained DenseNet121 [23] on spatial transcriptomic data using adaptive momentum based optimizer [34] with a batch size of 32 and initial learning rate of 0.001. Moreover, to limit model overfitting, the model training was stopped early if performance over validation set did not increase across 5 consecutive epochs [36].

**Comparison with Cellular Composition and Morphological Features**

In order to understand the association between various types of cells in a given spot and the corresponding gene expression patterns, we used cellular composition (counts of neoplastic, epithelial, connective, inflammatory, and necrotic) and nuclei morphological features for predicting spot level gene expression profile [37]. The nuclei were segmented and classified using a nuclear segmentation and classification model called HoVer-Net [38] pretrained on PanNuke dataset [39], [40]. HoVer-Net cellular boundaries prediction were then used for computing cellular counts, and shape-based (major and minor axis length, and major-to-minor axis length ratio) and color-based (RGB channel-wise mean) morphological features for each nucleus. Patch level features were obtained by computing the mean and standard deviation of nuclei-level features. We trained XGboost [41], Random Forest [42], Multi-Layer Perceptron [43], and Ordinary least squared regressor (OLS) over these features, but OLS outperformed and we used their results for performance comparison.

## 3 Results and Discussion

### 3.1 Visual Results

In order to assess the quality of predictions of gene expression levels using the proposed approach, we show the results of spot-level predictions overlaid on top of various test images together with their true gene expression levels for a number of genes in Fig. 3. It can be observed that the predicted expression shows significant correlation with true expression for these images. For each image, the Pearson correlation coefficient for a given gene in shown in the figure. The correlation scores of predicted and true gene expression for both GNAS and ACTG1 is 0.82 whereas for FASN and ERBB2 the correlation coefficient is 0.64 It is important to mention here that the predicted expression is generated using spot level image information only, and the correlation is expected to improve further by averaging the predicted expression across neighboring spots.

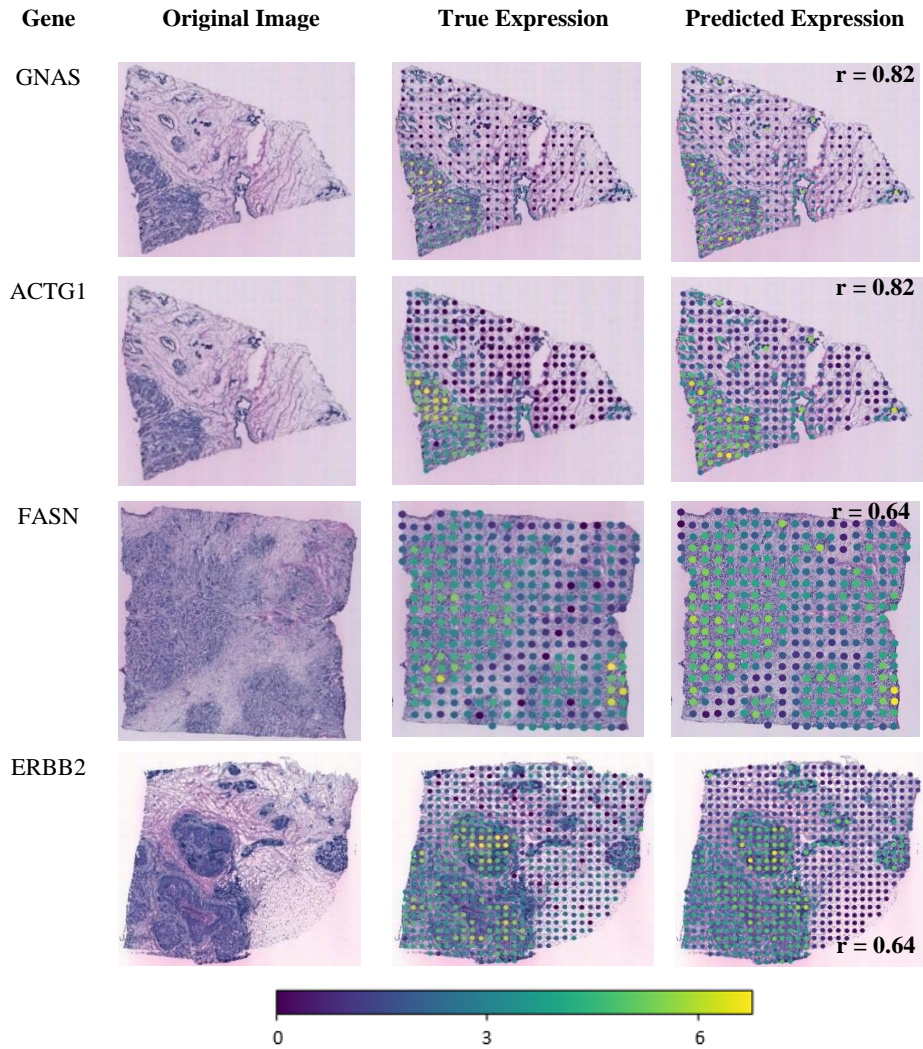

**Fig. 3.** True vs predicted gene expression. First column lists the names of selected genes; second columns show the original WSI consisting of multiple spots; third column shows a visualization of true spatial expression of a given gene at each spot; finally, the last column shows visualization of the predicted expression for a given gene at each spot. For two genes (GNAS and ACTG1), the Pearson correlation (r) is 0.82, while for two other genes FASN and ERBB2 the correlation is 0.64.

### 3.2 Quantitative Results

The median Pearson correlation coefficient between true and predicted gene expression levels for all 250 genes used in our analysis across all patients is shown in Fig. 4a together with their combined *p*-value. From the plot, it can be seen that the proposed NSL model has predicted the expression of 215 (out of 250) genes with a significant *p*-value. Moreover, the model has predicted 12-genes with a median Pearson correlation greater than 0.5. These genes include FASN, GNAS, ACTG1, ACTB, ERBB2, PSAP, TMSB10, PSMD3, PRDX1, EIF4G2, HSP90AB1, S100A11 and PFN1. Table-1 provides the median correlation coefficients of different genes along with the counts of genes whose expression was predicted with a high correlation score and significant p-values.

### 3.3 Gene set enrichment and pathway analysis

In order to understand the role of genes whose expression was predicted with high correlation using the proposed method, we performed gene set and pathway analysis using DAVID [44], [45]. Based on this analysis, we found that out of 215 genes with statistically significant correlation, 54, 56, 43, and 42 genes are respectively involved in cancer, pharmacogenomics, immune, and infections pathways [44], [45]. Among the genes whose expression was predicted with high (>0.5) correlation, ERBB2, FASN, GNAS, ACTB, and PSAP are considered as biomarkers for breast, gastric and prostate cancers. The most interesting aspect of our pathway analysis results in the context of our proposed approach is that genes whose expression was predicted with high correlation are involved in cell adhesion and tumor formation which is expected to have the most significant association on light absorption and hematoxylin binding in the tissue. This analysis clearly shows that the proposed approach of neural stain learning is able to learn tissue specific absorption characteristics and use them to predict gene expression levels in an effective manner.

### 3.4 Comparisons with DenseNet-121

Fig. 4(b) and Table-1 show the performance of DenseNet-121 in terms of *p*-value and median correlation score. From the plot, it can be seen that DenseNet-121 was able to predict the expression of only 170 genes with a significant *p*-value, but for the entire gene set the median Pearson correlation is less than 0.5.

### 3.5 Comparison with Cellular Composition and Morphological Features

Fig. 4(c) and Table-1 show the prediction accuracy of a model that uses cellular composition and morphological features for gene expression prediction. From the plot, it can be seen that the model was able to predict the expression of 209 genes with a significant p-value, however, only 4 genes are predicted with a median correlation coefficient greater than 0.5. This shows that NSL greatly outperforms other compared

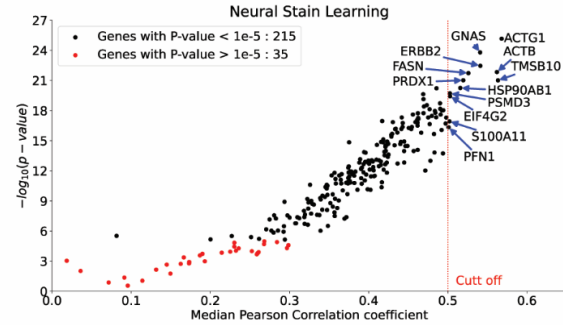

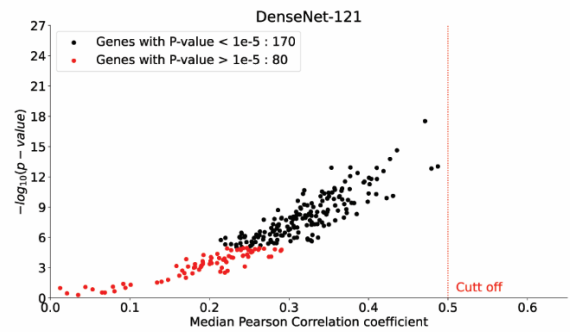

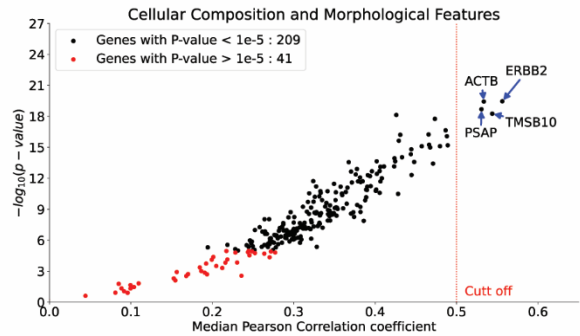

**Fig. 4.** Gene Expression vs *p*-value plots for (a) proposed neural stain learning model, (b) Dense-Net-121 and (c) Cellular Composition and Morphological Feature based regression. Black dots show genes predicted with significant (*p*-value < $10^{-5}$) while red dots show genes predicted with insignificant *p*-value. The vertical red line represents the cut-off threshold (genes predicted with correlation greater than 0.5).

**Table 1.** Median value of the correlation coefficient between predicted and true expression levels of selected genes and the total number of genes predicted with a median correlation score $r > 0.5$ and the total number of genes predicted with statistically significant correlation scores for different predictors.

| Gene / Method | GNAS | FASN | ACTG1 | ACTB | ERBB2 | PFN1 | # Genes with $r > 0.5$ | # Genes with $p < 10^{-5}$ |
|---|---|---|---|---|---|---|---|---|
| DenseNet | 0.42 | 0.40 | 0.36 | 0.42 | 0.47 | 0.29 | 0 | 170 |
| CC + MF | 0.48 | 0.43 | 0.44 | 0.53 | **0.55** | 0.48 | 4 | 209 |
| NSL | **0.54** | **0.52** | **0.58** | **0.56** | 0.54 | **0.50** | **12** | **215** |

approaches. Moreover, computing these features is laborious and computationally demanding.

## 4  Conclusions and Future Work

In this work, we investigated the contribution of stain information contained in histopathology images for image base gene expression prediction task in ST data. Handling the deficiency of previously proposed methods of not explicitly handling fundamental aspects of pathology images, i.e., histological staining, we proposed a novel neural stain deconvolution layer, which exploits tissue stain information for the gene expression prediction task. We have shown that for the gene expression prediction task, the proposed neural stain learning method significantly outperformed compared to methods using standard cellular features and deep learning based methods. Although morphometry may be informative for prediction, we have shown that gene expression of certain genes can be predicted using color information alone. Furthermore, the output of the proposed stain deconvolution layer can also be fed as input to a deep network. Apart from this, it can also be used for any other learning problem (classification, ranking, etc.) as long as a loss function can be formulated for it. We hope this study will open new ways of investigating the contribution of stain information to other computational pathology tasks, and assess stain layer performance when coupled with deep neural networks.

**Acknowledgments**
MD would like to acknowledge the PhD studentship support from GlaxoSmithKline. FM and NR supported by the PathLAKE digital pathology consortium which is funded from the Data to Early Diagnosis and Precision Medicine strand of the government's Industrial Strategy Challenge Fund, managed and delivered by UK Research and Innovation (UKRI).